# Ion irradiation-induced sinking of Ag nanocubes into substrates

Shiva Choupanian[1], Wolfhard Möller[2], Martin Seyring[3,6], Claudia Pacholski[4], Elke Wendler[1], Andreas Undisz[3,5], Carsten Ronning[1]

[1] Institute of Solid State Physics, Friedrich Schiller University Jena, Max-Wien-Platz 1, 07743 Jena, Germany

[2] Helmholtz-Zentrum Dresden-Rossendorf, Bautzner Landstraße 400, 01328 Dresden, Germany

[3] Otto Schott Institute of Materials Research, Friedrich Schiller University Jena, Löbdergraben 32, 07743 Jena, Germany

[4] Institute of Chemistry, University of Potsdam, Karl-Liebknecht-Str. 24-25, 14476 Potsdam, Germany

[5] Institute of Materials Science and Engineering, Chemnitz University of Technology, Erfenschlager Str. 73, 09125 Chemnitz, Germany

[6] Faculty of Electrical Engineering, Schmalkalden University of Applied Science, Blechhammer 9, 98574 Schmalkalden, Germany

**Abstract**

Ion irradiation can cause burrowing of nanoparticles in substrates, strongly depending on the material properties and irradiation parameters. In this study, we demonstrate that the sinking process can be accomplished with ion irradiation of cube-shaped Ag nanoparticles on top of silicon; how ion channeling affects the sinking rate; and underline the importance of the amorphous state of the substrate upon ion irradiation. Based on our experimental findings, the sinking process is described as being driven by capillary forces enabled by ion-induced plastic flow of the substrate.

**Corresponding authors**

shiva.choupanian@uni-jena.de, carsten.ronning@uni-jena.de

**Introduction**

Ion irradiation is today widely used for various purposes, e.g., in semiconductor industry for manufacturing processors, as this technique allows precise incorporation of any impurity into any (semiconductor) material. This precision originates from the knowledge gained over the past decades via intense research on ion-solid-interactions. This includes the capability to describe correctly the electronic and nuclear stopping mechanisms[1–3], damage formation[4,5], channeling of ions[6–14], sputtering of surfaces[15,16] as well as ion beam mixing at interfaces[17–21]. Many appropriate simulation tools are thus available for ion irradiation of bulk objects and thin films[22–26].

Today, being in an era of nanotechnology, the irradiated object sizes are often comparable to the projected ion range, which led to the discovery of many new effects, as the situation is different to the ion irradiation of bulk or thin films. A reduced dopant incorporation[27], an enhanced sputtering yield[28–35] as well as an enhanced dynamic annealing and thus lower damage formation[36–38] are a few of those effects, which recently opened the research field of 'ion-nanostructure-interactions'.

It has also been discovered that ion irradiation of a system consisting of metallic nanoparticles deposited on top of specific substrates can result in the sinking of the particles into the substrate with increasing ion fluence[39–43]. This effect has been described as an interplay of several effects. Ion irradiation of the substrate causes density changes and broken bonds. Plastic flow may release the resulting stress, such as by cooperative motion of defects[44,45]. Therefore, the irradiated substrate may be described as a fluid with a given viscosity[46]. Additionally, sputtering at the surface and ion beam mixing at the interface have to be considered, and finally, capillary forces may drive the burrowing of the nanoparticles into the substrate[39,40].

However, the ion-induced sinking of nanoparticles has been investigated up to date only for a limited combination of materials so that its understanding must be considered as rudimentary. The significance of the individual effects and their interplay is not clear so far. Furthermore, the influence of the material chemistry in general, and in particular its influence on the ion beam mixing process at the nanoparticle-substrate interface, is also unclear. This also holds for any channeling effects and the role of the phase (crystalline/amorphous) of the substrate.

In this study, we investigate the sinking behavior of single crystalline Ag nanocubes into different substrates upon ion irradiation. We compare our results with existing literature data for Au nanoparticles and also discuss the sinking mechanism with respect to other processes leading to burrowing, such as metal-assisted etching[47–49] or thermal oxidation[50,51]. Furthermore, irradiation experiments have been performed in random and channeling directions with respect to the crystal axis of the Ag nanocubes in order to identify the role of the energy loss on the sinking process. Finally, we also varied the irradiation temperature for the case of silicon substrates, as Si is amorphized at room temperature upon ion irradiation; whereas it remains crystalline at elevated temperatures.

**Silicon substrates and room temperature irradiations**

Single crystalline silver (Ag) nanocubes of 100 nm size exhibiting {001} side planes were dispersed on silicon <100> substrates via spin coating. After this process, all Ag nanocubes are oriented normally on the substrates with the <100> axis being aligned with the substrate surface normal[52]. These samples were then irradiated with 300 keV Ar$^+$ ions in subsequent steps with increasing ion fluence in two alternative directions, either aligned or random (tilted by 7°) in respect to the crystallographic axis of both the Ag nanocubes and Si substrate. As calculated using SRIM[23], the flat-surface distributions of recoil atom generation, which determines surface sputtering and atomic mixing at interfaces, peak at ~ 250 nm and ~70 nm in Si and Si covered with 100 nm Ag ion respectively, being irradiated in random direction (see the supplementary information for details, figure S1). Note that the mean depths are significantly larger for the aligned channeling direction.

The top row of **figure 1** shows scanning electron microscopy (SEM) images of cross-sectioned Ag nanocubes, which were irradiated along the channeling direction. It can be clearly seen that the Ag nanocubes sink deeper and deeper into the substrate with increasing ion fluence. Thus, this ion induced burrowing effect does also occur for cube-shape particles as well as for the Ag-Si material combination, and not only for spherical Au or Pt nanoparticles on $SiO_2$, as reported in references [39,40]. Furthermore, it is evident that the nanocubes' side surfaces are enclosed by the Si substrate already after the first low fluence irradiation (see yellow dotted lines in figure 1 showing the Si surface). This clearly indicates the action of capillary forces.

The edges of the Ag nanocubes become rounded off, and the overall size decreases with increasing ion fluence, which both can be attributed to sputtering[52]. One can further observe that small clusters are formed at the interface between the Ag nanocubes and the silicon substrate for ion fluences larger than 1.2 x 10$^{16}$ ions/cm$^2$. These small clusters originate from ion beam mixing and subsequent phase separation at the interface of the two non-solvable Ag/Si components[53,54], and grow in size with increasing ion fluence, likely affected by Ostwald ripening. This will be discussed in further detail below.

The bottom row of figure 1 shows the direct comparison with the system under irradiation in random direction. We again observe that the Ag nanocubes sink more and more into the substrate at increasing ion fluence, but there are some distinct differences compared to the aligned channeling irradiation. It is evident that the edge rounding is more efficient at random irradiation. Under aligned irradiation, ion channeling allows the ions to penetrate deeper into a crystal, and simultaneously results in a less dense collision cascade and less transferred energy to the target atoms thus leading to a reduced sputtering yield. The different sputtering yields are also recognizable by comparing the width of the Ag nanocubes after irradiation with ion fluences of 3.2 × 10$^{16}$ and 4 × 10$^{16}$ cm$^{-2}$. The Ag nanocubes, which received aligned irradiation, are considerably wider with a width of ∼ 90 nm compared to ∼ 80 nm for random irradiated nanocubes at the highest ion fluence. Also, their height remains larger to some extent.

The observed morphology changes have been compared with TRI3DYN simulations, which is a Monte Carlo program tool describing ion irradiation effects in three-dimensional amorphous multicomponent material systems using the binary collision approximation (BCA) (see discussion in supplementary information & figure S2). The experimentally observed sinking into the silicon substrate is not reflected by such simulations, as they are of purely collisional nature and neglect any collective effects governed by thermodynamic and/or chemical properties under the influence of energetic ion bombardment, to which the sinking has to be attributed.

All samples in figure 1 have been additionally analyzed with Rutherford backscattering spectroscopy RBS/channeling to investigate the integral ion beam-induced damage in the crystal structure of the Ag nanocubes (for data see supplementary information, figure S3). Non-irradiated Ag nanocubes on silicon show a minimum yield of around 0.2 for both the Ag and Si signals indicating high crystallinity of both constituents. The minimum yield of the Si signal becomes 1.0 immediately after the first Ar$^+$ irradiation with an ion fluence of 4 × 10$^{15}$ cm$^{-2}$ demonstrating the low amorphization threshold of silicon at room temperature, which is in agreement with references[55,56]. The minimum yield of the Ag signal increases with increasing ion fluence until it reaches also values of 0.8-1.0 but after ion fluences of 2-4 × 10$^{16}$ cm$^{-2}$. This verifies that metals have a high resistance against amorphization [57], but point and extended defect formation is occurring at such high fluences[58]. Comparing the minimum yields, we further find that defect formation in the Ag nanocubes starts in channeling direction at higher ion fluences compared to the random irradiation.

In order to corroborate the findings of the integral RBS/channeling analysis, cross-section lamellas have been prepared and the irradiated Ag/Si system was examined with regard to their defect structure using high-resolution transmission electron microscopy (HRTEM). **Figure 2** illustrates TEM images of two Ag nanocubes, which were irradiated with 300 keV Ar$^+$ ions at an ion fluence of 1.6 × 10$^{16}$ ions/cm$^2$ in random (top row) and aligned (bottom row) direction. Strong and bright spot-like/spotted contrast can be seen in the lower magnified TEM images (left column). Such diffraction contrast appear in the vicinity of areas where the lattice is distorted, therefore this contrast is the indication of ion beam-induced crystal defects and defect complexes in the crystal structure of the Ag nanocubes. The main difference in the lower magnified images is less contrast in the Ag nanocube irradiated in channeling direction compared to random direction. This agrees with the RBS results that ion irradiation in channeling direction produces less defects, caused by a less dense collision cascade. However, the HRTEM micrographs with higher magnification in figure 2 depicting the center as well as interfaces of these particular Ag nanocubes clearly show that the Ag nanocubes remain essentially single crystalline, as the respective fast Fourier transforms (FFT, insets) that resemble diffractograms show the same maxima and thus the same orientation. Furthermore, the lattice spacings and maxima in the FFT fit with the fcc crystal structure of Ag. For more TEM images of various interfaces see supplementary information (figures S4 & S5).

Another set of HRTEM images was taken from an Ag nanocube, which was irradiated with 2.5 times the ion fluence ($4 \times 10^{16}$ ion/cm$^2$) in random direction, being displayed in **figure 3**. Compared to figure 2, it is evident that the size of the spot-like/spotted contrast in the lower magnified image is larger, which means that the extended defects and defect clusters grew throughout the entire Ag nanocube. The HRTEM images of the Ag nanocube, along with the more diffuse maxima in the respective FFTs, corroborate the presence of a larger number of defects in the crystal structure, but it still exhibits straight lattice fringes. This is further visualized by applying Fourier filtering on an area of $5 \times 5$ nm$^2$ at the center of the irradiated Ag nanocube as displayed in **figure** 3 (bottom, right). Furthermore, one can clearly observe that the shape of the Ag nanocube changed and became rounder. This can be assigned to the continuing ion beam irradiation, resulting in mixing, sputtering and re-deposition, eventually causing phase separation in the vicinity of the interface of the Ag nanocube/Si substrate, and the tendency for minimizing the interfacial free energy.

As we have already observed in figure 1, a halo of small clusters is formed in the vicinity of the interface. A HRTEM image of such a cluster with a diameter of about 5 nm is depicted in figure 3 (bottom, left). These clusters are also crystalline but do not exhibit the same orientation as the Ag nanocube. The interatomic distance in the cluster has been determined to be 0.243 nm, which again fits the fcc structure of silver. Additional electron dispersive X-ray (EDX) analysis confirmed that these are indeed Ag clusters (see supplementary information, figure S5), which are formed by ion beam mixing of the Ag/Si interface and subsequent phase separation of the two non-solvable constituents at room temperature[59].

**Silicon substrates and irradiation at 300 °C**

It has been shown that ion irradiation of silicon at high temperatures prevents its amorphization, as created defects recover immediately due to the kinetic activation and enhanced diffusion at those elevated temperatures[60]. We confirmed this for our irradiation parameters using a bare Si substrate and performing RBS/channeling analysis afterward (see supplementary information, figure S7).

In **figure 4** we compare two SEM images of cross-sectioned Ag nanocubes on silicon irradiated in random direction with an ion fluence of $4 \times 10^{16}$ ion/cm$^2$ at room temperature and at an elevated temperature of 300 °C. It is evident that the irradiation at higher temperature does not result in any sinking of the Ag nanocube into the silicon substrate. Although sinking is not observed at this elevated temperature, the wetting angle of the nanocube has been slightly modified. This effect is comparable to the heat treatment of Au nanoparticles on SiO$_2$[51] in which evaporation of gold and dewetting is directly followed by ridge formation of SiO$_2$ at the triple line, to create an air phase angle. Furthermore, it is apparent that sputtering seems to be more significant for the high-temperature irradiation, as the Ag nanocube is never (partly) buried. However, the nanocube experiences a noticeable increase in width at 300°C, in contrast to RT. Finally, one can observe that the morphology fully changes and that only some small clusters at the Ag-Si interface appear. This is again due to the ion beam-induced mixing and phase separation processes, but at higher temperatures the influence of diffusion becomes more pronounced. The diffusion coefficient of Ag in Si as well as on its surface significantly increases when the temperature rises to 300 °C[61,62].

**Zinc oxide substrates and room temperature irradiation**

Zinc oxide (ZnO) is a material with a high radiation hardness, which even does not get amorphized at room temperature upon our irradiation parameters[5]. We dispersed Ag nanocubes on a ZnO substrate in the same way as for the silicon substrates and irradiated them with 300 keV Ar$^+$ ions in aligned direction at room temperature with an ion fluence of $2 \times 10^{16}$ ion/cm$^2$. The SEM images in **figure 5** compare cross-sectional views of such an Ag-nanocube with one on top of silicon, that were exposed to the same irradiation condition. The Ag nanocube does not sink into the crystalline ZnO but is sinking into the amorphous Si substrate under these conditions. In figure 5c the TRI3DYN simulation result of an Ag nanocube on Si irradiated at the same ion fluence is displayed. The shape and size of the simulated nanocubes resemble well the Ag nanocube on ZnO.

**Discussion and modelling**

A quantitative evaluation of the SEM images shown in figure 1 is displayed in **figure 6** analyzing at least 15 cross-sectioned nanocubes for each ion fluence using *ImageJ*[63]. Let us first consider the range of fluences below ~$2 \times 10^{16}$ ions/cm$^2$. Within the statistics, there is no clear trend in the development of the total height (figure 6a) and width (figure 6b) of the nanocubes with ion fluence. However, one might delineate some reduction of the total height at a more or less constant width. This would indicate that material loss from the nanocubes primarily occurs by sputtering from the top and ion-induced transport at the bottom. From the recoil generation distribution (figure S1), one may expect that both channels contribute to a similar extent. Surprisingly at the first glance, the development of both width and height is essentially independent of the alignment. However, under aligned compared to random irradiation, the recoil generation distribution moves into depth to some extent, but also decrease in amplitude, so that the amount of forward transported atoms at the bottom is little influenced by the orientation. However, recoil atom generation will be enhanced in the underlying Si bulk under aligned irradiation, which promotes the sinking rate as confirmed by figure 6c. Nevertheless, also defects formed within the nanocubes' crystal structure, as identified by the integral RBS analysis and TEM investigations, and/or recoiled atoms from the surrounding substrate, which is amorphized in the initial stage, might contribute to the observations and complicate their interpretation.

The trend of the plots is changing for ion fluences above $\sim 2 \times 10^{16}$ ions/cm$^2$. As seen in figure 6c, the nanocubes get more and more buried, to which also the redeposition of silicon atoms, which are sputtered from the flat surface in the vicinity of the nanocubes, may contribute. The total height and width of the nanocubes (figure 6a,b) remain almost constant for the aligned direction, whereas for random irradiation the width is reduced. This can be assigned to the reduced lateral straggling in aligned direction, resulting in a narrower collision cascade. Therefore, it is confirmed that channeling is still effective though to a lower extent. The rates of sinking for both alignments become comparable. This is attributed to the progressive prevention of channeling due to defect formation within the nanocubes, which leads to comparable energy deposition in the Si substrate and consequently similar sinking rates for both orientations (**figure** 6c) regardless of the irradiation direction.

From the ion irradiation of the Ag nanocubes on Si at elevated temperature and Ag nanocubes on ZnO at RT, it is evident that the sinking of the Ag nanocubes only occurs upon ion irradiation if the substrate is amorphous and not crystalline. This is in agreement with literature, where such sinking has been observed for Au and Pt on SiO$_2$ and not for Au on sapphire[39,40].

Crystalline materials do not allow strong density changes upon ion irradiation, especially not in a steady-state situation when defect initiations and recovery are in balance due to the fast diffusion of point defects, such as vacancies and/or interstitials. On the other hand, significant density changes can occur in amorphous materials resulting in more or less dense material [64]. Plastic deformation and plastic flow are mechanisms to relieve such density changes/stress in ion-irradiated regions[45,65–67]. This has been recently identified for ion beam induced ripple formation[68] and also as one of the driving boundary conditions for the sinking process of metallic nanoparticles into amorphous substrates upon ion irradiation[39,40]. Each ion impact causes movements of substrate atoms in its cascade volume. After the dissipation of the energy, the situation is frozen until the next ion hits the same volume. Thus, continued irradiation results in overlapping cascade volumes, and the sum of many ion impacts results in a dynamic situation, where one can describe the solid as a Newtonian fluid with a given viscosity[45,65,69]. However, this is not the driving force for the sinking process. The driving force are capillary forces given by the respective surface and interface energies of the involved phases/materials.

A detailed view to figure 1, but also much more obvious in the low magnified TEM images of figure 2, clearly manifests the formation of a meniscus (or in other words wetting) of the silicon substrate on the Ag nanocube's side surfaces. Thus indeed, one can describe the sinking process upon irradiation such as a sinking solid (here the crystalline Ag nanocubes, which do not undergo amorphization) into a fluid (the amorphous silicon) driven by capillary forces. Note, gravitation can be fully ruled out, as the samples are mounted vertically during the irradiation process.

A quantitative description of the sinking by capillary forces has been introduced by Hu et al.[39] and further developed by additionally considering enhanced sputtering by Klimmer et al.[40]. We applied this model to our experimental data (see supplementary information for details), even though the model considers spherical particles. Figure 6d replots the data of figure 6a&c and depicts the height of the Ag nanocubes above the Si substrate as a function of ion fluence for the two irradiation directions, which has been fitted with the model (lines). The trend of the experimental data can again be divided to the ion fluences below and above $2 \times 10^{16}$ ions/cm$^2$. Similar to figure 6c, it can be observed that the aligned irradiated nanocubes sink with a faster rate into the viscous Si at low ion fluences; whereas, the sinking rates are comparable for high ion fluences due to the minor channeling effects at these fluences.

From the model reasonable parameters could be extracted, and we determined ion-induced viscosity values of about $3.39 \times 10^{23}$ Pa*ions/cm$^2$ and $1.17 \times 10^{24}$ Pa*ions/cm$^2$ for the aligned and random irradiation directions, respectively. These numbers illustrate how the viscosity changes; where lower numbers indicate lower viscosity. A value of $2 \times 10^{24}$ Pa*ion/cm$^2$ has been reported for 2 MeV Xe$^+$ irradiation of Si in random direction [44], in good agreement with our experiment. On the other hand, our value for the aligned irradiation is close to the value of $\sim 1.5 \times 10^{23}$ Pa*ion/cm$^2$ reported for Xe irradiations of SiO$_2$ substrates[45,70], and it has been shown that ion induced viscosity correlates inversely with nuclear stopping density[70]. This fits with the observation: when irradiation is along the channeling direction, less energy is transferred to the nanocubes and most of the nuclear stopping energy is deposited to the Si substrate. Thus, higher plastic flow is induced in the Si substrate, e.g. it has lower viscosity. Accordingly, nanocubes irradiated aligned to the channeling direction sink faster into the substrate. We also determined that the constant 'k', which links the sputter yield to the size of the nanoparticles, is an order of magnitude higher for randomly irradiated nanocubes. This finding is in line with the trend observed in figure 6d for ion fluences exceeding $2 \times 10^{16}$ ions/cm$^2$.

**Comparison to thermal and chemical induced sinking processes**

As described earlier, silver nanocubes can sink into crystalline (amorphizable) Si by ion irradiation. Mere heat treatment of such samples is not sufficient for nanoparticles to penetrate into crystalline Si substrates. However, a movement of metallic nanoparticles into natural or thermally grown $SiO_2$ layers on crystalline Si was observed during a heat treatment. This sinking process results in the formation of a pore in the substrate along the path of the particles.

Two different explanations for the pore formation have been proposed. Ono et al. suggested that the $SiO_2$ might react with metallic nanoparticles at elevated temperatures [48]. In this case, the metallic nanoparticles should act as a mediator for the decomposition of $SiO_2$, leading to the formation of pores in the $SiO_2$ layer [48]. The pore depth then depends on the thermal annealing time. In contrast to this model, Bowker et al. attributed the pore formation to different growth rates for a $SiO_2$ layer on the crystalline silicon during thermal annealing [50]. Underneath the metallic nanoparticles, the growth of the $SiO_2$ layer should be hindered due to the limited access of oxygen to the silicon surface below the nanoparticles.

Other explanations for the pore formation have been given for similar samples consisting of gold nanoparticles on amorphous $SiO_2$ substrates treated at high temperatures of at least 900 °C [51,71,72]. The nanoparticles sink progressively into the $SiO_2$ substrate with heating time suggesting a diffusion-controlled process [51], which differs significantly from the results obtained by ion irradiation. Here, the nanoparticle burrowing continues until the nanoparticles are embedded in the substrate matrix, and further ion irradiation would not lead to deeper immersion into the substrate, because capillary forces then vanish. In addition, ion irradiation of the system leads only to the amorphization of the Si substrate up to the ion range, which also limits the sinking depth.

The partial transformation of crystalline silicon into a $SiO_2$ layer under metallic nanoparticles has been reported for metal-assisted chemical etching of Si [49]. In this case, crystalline Si substrates covered with metallic nanostructures are immersed in a solution consisting of hydrofluoric acid and an oxidizing agent such as hydrogen peroxide. The presence of the metallic nanoparticles catalyzes a redox reaction that leads to the formation of pores [49,73]. In this case, the pore depth can be controlled by the reaction time. Depending on the reaction conditions, the Si underneath the nanoparticles can either be converted directly into soluble chemical compounds such as silicon tetrafluoride or silicon hexafluoride, or first into $SiO_2$, which is subsequently dissolved by the hydrofluoric acid.

Our results presented here are more similar to the interaction of metallic nanoparticles on amorphous $SiO_2$ substrates treated at high temperatures of at least 900 °C. De Vreede at al. suggested that the reduction of the interfacial free energies at the air/nanoparticle/substrate triple line leads to the formation of pores created by the sinking of gold nanoparticles into amorphous $SiO_2$ [51]. In this case, the gold evaporates during heating and deep pores are formed until all the gold has disappeared. The process was explained by the occurrence of capillary forces, as we proposed for the sinking of silver nanocubes into Si by ion irradiation. This model was further developed by Tregouet et al. who proposed diffusion of gold into a thin liquid silica layer at the interface between the two materials [71]. The resulting diffusiophoretic effect is then the cause of pore formation. More recently, a lowering of the glass transition temperature of $SiO_2$ due to a diffusion of gold into this material was also held responsible for the pore formation by Gosavi et al.[72].

Moreover, ion irradiation leads to morphological changes in the nanoparticles due to sputtering. Thermal annealing can also result in morphology changes, but these are based on the minimization of the surface energy of the nanoparticles [74,75].

*Conclusion*

We have demonstrated that cube-shaped metallic nanoparticles can sink into substrates upon ion irradiation, just like spherical nanoparticles shown in literature, which strongly indicates that this is a nanoparticle shape-independent process. Furthermore, the silver/silicon system is another material combination that shows sinking, suggesting that a lot of (noble) metals should be applicable for harvesting this effect. The main reason for this phenomenon seems that most metals do not amorphize upon ion irradiation, and another necessary condition is likely that the metal and substrate must be immiscible. Additionally, our channeling experiments clearly show that the irradiation of the substrate is essential for the burrowing effect rather than the irradiation of the nanoparticle, as the aligned irradiations showed faster dynamics. The substrates must be either amorphous or at least amorphizable upon low ion fluence irradiation.

The change of the morphology of the irradiated nanoparticle can be well described by TRI3DYN simulations, up to the point until the sinking/burrowing effect becomes significant. This underlines that collective and thermodynamically driven effects in the substrate are indeed responsible for the sinking process: ion irradiation causes amorphization with significant density changes associated with mechanical stress, and plastic deformation and flow are mechanisms to relieve such stress. Each ion impact causes movements of substrate atoms in its cascade volume. The continued irradiation results in overlapping cascade volumes, and the sum of many ion impacts results in a dynamic situation, where one can describe the solid as a Newtonian fluid with a given viscosity. Then capillary forces result in sinking of the nanoparticle into the substrate, comparable to the sinking of nanoparticles into $SiO_2$ upon a heat treatment, which can be also quantitatively modelled. The system tends to reach an equilibrium state to eliminate the imbalance of the interfacial free energies at the triple line of air/nanoparticle/substrate until the nanoparticle is fully covered.

**Acknowledgments**

We thank the Deutsche Forschungsgemeinschaft (DFG) for financial support through the project "Energy induced nanoparticle substrate interactions" (Ro1198/22-1 and PA925/6-1) as well as for the financial support for the HR-TEM via the Major Research Instrumentation Program (Inst 275/391-1).

**Experimental details**

**Sample preparation.** Single crystalline silver (Ag) nanocubes of 100 nm size dissolved in ethanol and PVP 55 kDa (Sigma-Aldrich, Eschenstr. 5, 82024 Taufkirchen, Germany) owing *{001} side planes were* dispersed on Si <100> and ZnO <0001> substrates via spin coating. All the single crystalline Ag nanocubes are then oriented with their <100> axes being parallel to the substrate normal, which is a necessary condition for all channelling experiments.

**Ion irradiation.** Various samples have been irradiated with 300 keV $Ar^+$ ions in subsequent steps with ion fluences ranging from $8 \times 10^{15}$ to $4 \times 10^{16}$ ions/cm$^2$, either aligned or random to the channeling direction at room temperature (RT) or 300 °C.

***Rutherford backscattering analysis*** *(RBS/channeling) measurements were performed with* 1.4 MeV $He^+$ ions and a beam spot of about 1 mm in diameter, thus, characterizing an ensemble of Ag nanocubes. These RBS measurements were also done either in aligned *channeling (0° - ion beam parallel to the surface normal) or random (7°) geometry.*

**Cross-section and lamella preparation.** Cross-section lamellas have been prepared using a FEI Helios 600i NanoLab focused ion beam system. A selected area of 5 x 12 µm$^2$ has been covered with ~ 1 µm Pt to preserve the surface from $Ga^+$ ion damage. Platinum was deposited in two stages. First deposited using an electron beam and next using a $Ga^+$ ion beam, respectively. The thicknesses of the electron deposited Pt layers were different for various samples. For cross-sectioning, the covered area is etched using a low energy $Ga^+$ ion beam. In order to prepare lamellas, the covered area is etched from both sides, step by step down to the desired thickness. The thickness of the lamella was lower than 100 nm in order to be able to investigate the nanocubes from the both sides of the lamella. Then the lamella was transferred to a lamella grid.

**Scanning electron microscopy** (SEM) images from the cross-sectioned Ag nanocubes after ion irradiation were also acquired using the same FEI Helios 600i NanoLab focused ion beam system.

**Image analysis.** Each SEM image from the cross-sectioned Ag nanocube has been analyzed with the image analysis software *ImageJ* [63] to estimate the fraction/height of each Ag nanocube sunk into the substrate and remained on top of the substrate.

**High-resolution transmission electron microscopy** *(HRTEM)* images were taken using a $C_S$-corrected JEOL NEOARM 200F. Lamellas were prepared out of the irradiated samples using a focused ion beam system Helios 600i NanoLab. Fast Fourier transforms were acquired from the HRTEM images using the digital micrograph software GMS3 (Gatan). The elemental composition was determined via **energy dispersive X-ray spectroscopy (EDS)** using a Double SDD (JEOL Centurio), as shown in the supplementary information.

[52] S. Choupanian, W. Möller, M. Seyring, C. Ronning, Low-energy ion channeling in nanocubes, Nano Res. 16 (2023) 1522–1526. https://doi.org/10.1007/s12274-022-4723-6.

[53] B. Satpati, P.V. Satyam, T. Som, B.N. Dev, Nanoscale ion-beam mixing in Au–Si and Ag–Si eutectic systems, Appl. Phys. A. 79 (2004) 447–451. https://doi.org/10.1007/s00339-004-2703-1.

[54] G.C. Rizza, M. Strobel, K.H. Heinig, H. Bernas, Ion irradiation of gold inclusions in SiO2: Experimental evidence for inverse Ostwald ripening, Nucl. Instruments Methods Phys. Res. Sect. B Beam Interact. with Mater. Atoms. 178 (2001) 78–83. https://doi.org/10.1016/S0168-583X(01)00496-7.

[55] S. Prussin, D.I. Margolese, R.N. Tauber, Formation of amorphous layers by ion implantation, J. Appl. Phys. 57 (1985) 180–185. https://doi.org/10.1063/1.334840.

[56] J.R. Dennis, E.B. Hale, Crystalline to amorphous transformation in ion-implanted silicon: a composite model, J. Appl. Phys. 49 (1978) 1119–1127. https://doi.org/10.1063/1.325049.

[57] K. Nordlund, M. Ghaly, R.S. Averback, M. Caturla, T. Diaz de la Rubia, J. Tarus, Defect production in collision cascades in elemental semiconductors and fcc metals, Phys. Rev. B. 57 (1998) 7556–7570. https://doi.org/10.1103/PhysRevB.57.7556.

[58] B. Johannessen, P. Kluth, D.J. Llewellyn, G.J. Foran, D.J. Cookson, M.C. Ridgway, Amorphization of embedded Cu nanocrystals by ion irradiation, Appl. Phys. Lett. 90 (2007) 073119. https://doi.org/10.1063/1.2644413.

[59] K.H. Heinig, T. Müller, B. Schmidt, M. Strobel, W. Müller, Interfaces under ion irradiation: growth and taming of nanostructures, Appl. Phys. A Mater. Sci. Process. 77 (2003) 17–25. https://doi.org/10.1007/s00339-002-2061-9.

[60] L. Pelaz, L.A. Marqués, J. Barbolla, Ion-beam-induced amorphization and recrystallization in silicon, J. Appl. Phys. 96 (2004) 5947–5976. https://doi.org/10.1063/1.1808484.

[61] T.C. Nason, G. -R. Yang, K. -H. Park, T. -M. Lu, Study of silver diffusion into Si(111) and SiO 2 at moderate temperatures, J. Appl. Phys. 70 (1991) 1392–1396. https://doi.org/10.1063/1.349547.

[62] L. Chen, Y. Zeng, P. Nyugen, T.L. Alford, Silver diffusion and defect formation in Si (1 1 1) substrate at elevated temperatures, Mater. Chem. Phys. 76 (2002) 224–227.

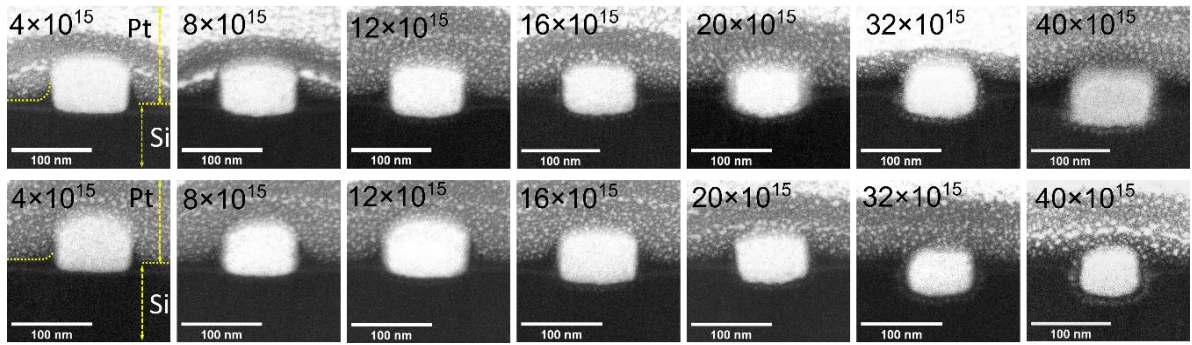

**Figure 1**: High-resolution scanning electron microscopy (SEM) images from cross-sectioned Ag nanocubes irradiated with 300 keV Ar+ ion at RT at the indicated ion fluence. (Top row) The ion beam was aligned with respect to the channeling direction of the Ag cubes and Si substrates, thus parallel to the substrate normal. (Bottom row) Tilted irradiation with 7° with respect to the substrate normal, utilizing a random direction. The ion fluences are expressed in units of ion/cm$^2$. Dotted lines are guides to the eye and follow the Si surface. Platinum was used as protective layer upon cross-section preparation of the specimen.

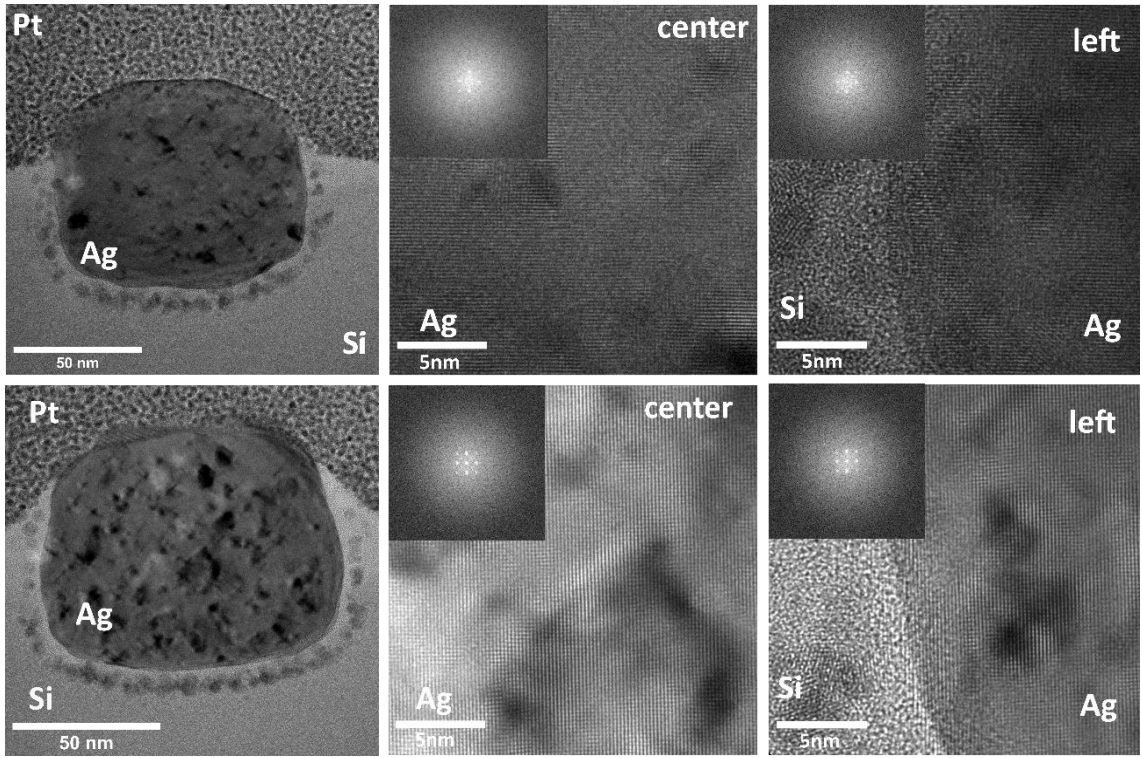

**Figure 2**: Transmission electron microscopy (TEM) images of Ag nanocubes sunken into the silicon substrate, after irradiation with 300 keV Ar$^+$ ions at the fluence of $1.6 \times 10^{16}$ ion/cm$^2$. (Top row) Irradiated in random direction. (Bottom row) Irradiated in aligned direction. The overview images are shown in the left column. The high-resolution images were taken in the center as well as at the interfaces of each Ag nanocube to the Si substrate (here the interface on the lefthand side of each particle is shown). The insets show respective FFTs taken from $10 \times 10$ nm$^2$ areas, resembling local diffractograms.

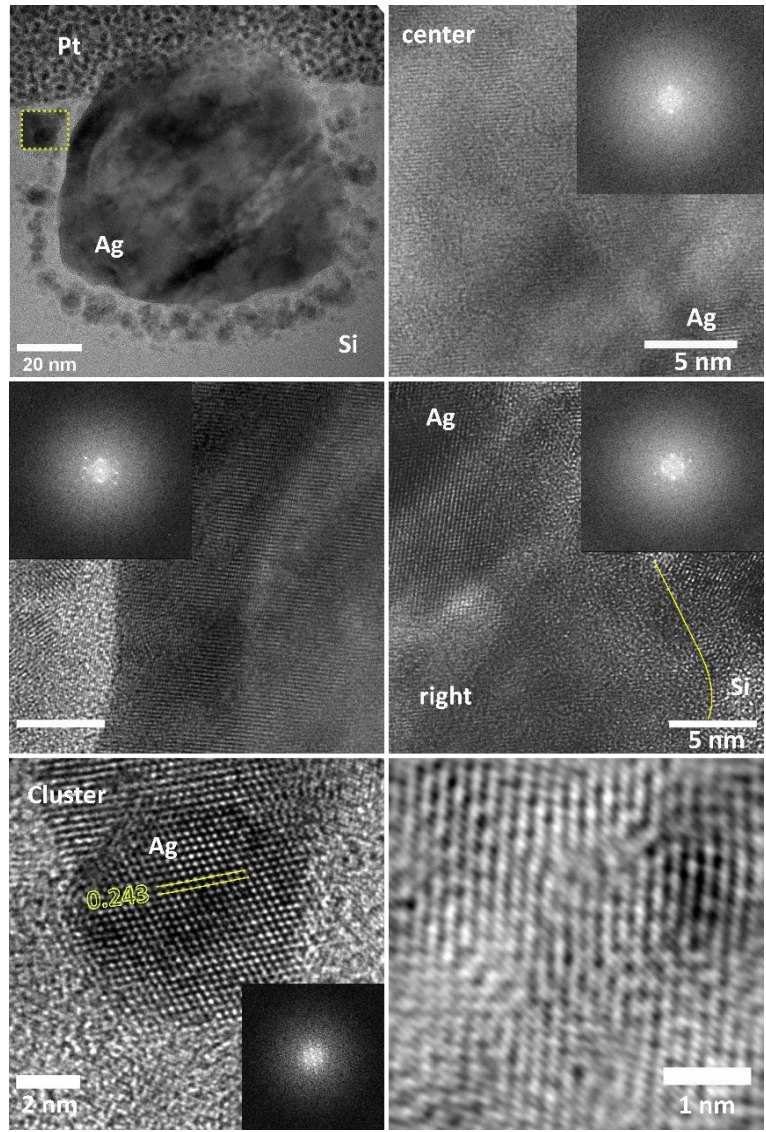

**Figure 3:** Transmission electron microscopy (TEM) images of an Ag nanocube, which was irradiated with 300 keV Ar+ ions at the influence of 4 × 10$^{16}$ ion/cm$^2$ in random direction on top of a silicon substrate. A low-resolution overview image is shown in the top row on the left side. The other three high-resolution images were taken in the center as well as at the right and left interfaces. The insets show respective Fourier diffractograms taken from 10 × 10 nm$^2$ areas. (Bottom, left) High-resolution image of a cluster, which is marked in the overview image. (Bottom, right) The Fourier filtered image was calculated/processed from an area of 5 × 5 nm$^2$ from the center of the nanocube.

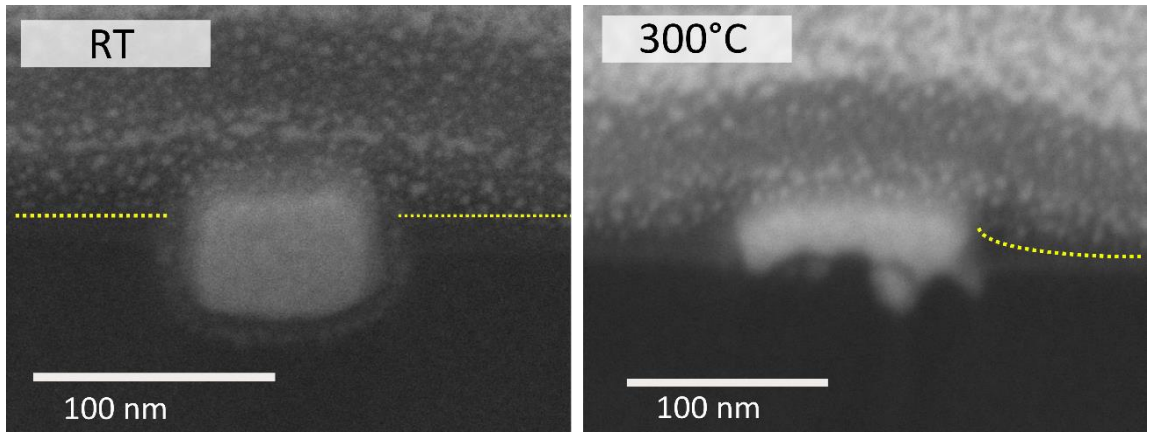

**Figure 4:** Scanning electron microscopy (SEM) images of cross-sectioned Ag nanocubes irradiated with 300 keV Ar⁺ ions with an ion fluence of 4 × 10$^{16}$ ion/cm$^2$ at (a) room temperature and (b) 300 °C.

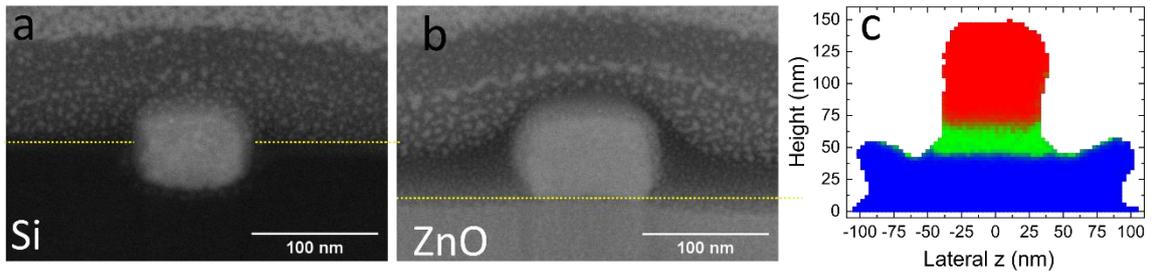

**Figure 5:** Scanning electron microscopy (SEM) images form cross-sectioned Ag nanocubes on *(*a) silicon and (b) ZnO substrates irradiated with 300 keV Ar$^+$ ions at 2 × 10$^{16}$ ion/cm$^2$ in aligned direction. The sample (b) has been covered with a thin layer of carbon before cross section preparation. (c) TRI3DYN simulation result of an Ag nanocube on Si irradiated with an ion fluence of 2 × 10$^{16}$ ion/cm$^2$.

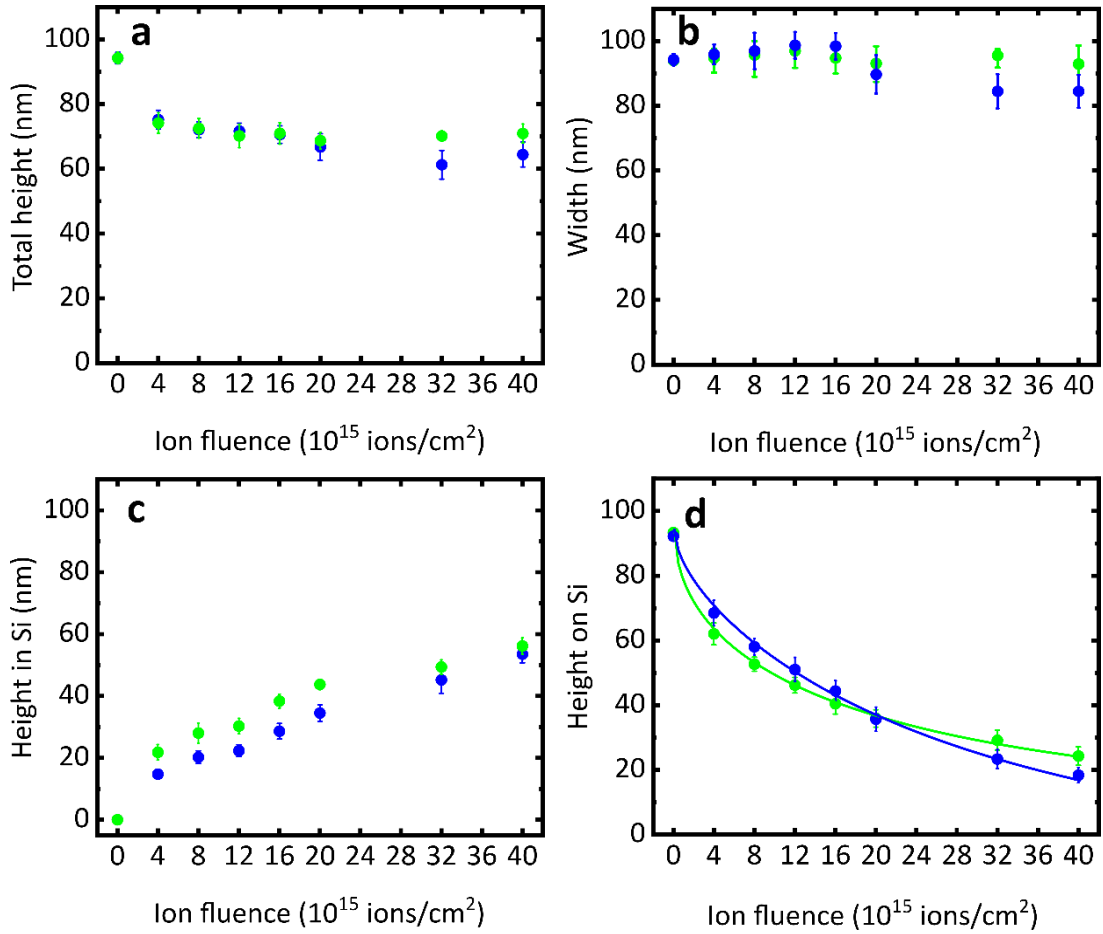

**Figure 6:** Sinking dynamics of Ag nanocubes in silicon irradiated in random (blue) and aligned (green) direction as a function of ion fluence. (a) Total height of the nanocubes as a function of ion fluence. (b) Width of the nanocubes as a function of ion fluence. (c) Bottom height of the nanocubes that sunk into the substrate. (d) Height of the nanocubes outside the silicon substrates. The circles represent the experimental data, whereas the lines are fits based on the model of Klimmer et al. [40].